# Identification of a replicable optical security element using laser speckle

A.M. SMOLOVICH[1*], A.V. FROLOV[1], L.D. KLEBANOV[2], I.D. LAKTAEV[1], A.P. ORLOV[1,3,4], P.A. SMOLOVICH[5], O.V. BUTOV[1,6]

[1]*Kotelnikov Institute of Radioengineering and Electronics of RAS, 125009 Moscow, Russia*
[2]*IPG Photonics Corp., Oxford, MA, USA*
[3]*Institute of Nanotechnology of Microelectronics of RAS, Moscow, Russia*
[4]*Biomedical Science & Technology Park, I.M. Sechenov First Moscow State Medical University, Moscow, Russia*
[5]*Twilio Spain SL, Madrid, Spain*
[6]*HSE University, Moscow, Russia*
*\*Corresponding author: asmolovich@petersmol.ru*

**ABSTRACT**

An optical security element containing an area of random rough relief is proposed. It combines the low cost of mass replication inherent in traditional security holograms with the impossibility of holographic copying, when the wave restored by the hologram is rewritten as a copy of this hologram. The proposed optical element is also protected from contact and photographic copying. Laboratory samples of optical elements were obtained by taking replicas of a rough surface. Identification of the authenticity of optical elements was demonstrated by calculating the cross-correlation of speckle patterns produced by coherent light scattered off different replicas. It is assumed that the proposed security elements can be mass-produced on standard equipment for embossing security holograms.

**Key words**: Optical security element, Authenticity identification, Random relief, Speckle, Correlation

1. ## Introduction

Optical protection systems are very diverse [1-4]. Among them, security holograms are the most common means of protection against counterfeiting of identity cards, financial and credit documents and other objects. First of all, they look attractive because of their simplicity and low cost of replication through embossing. In particular, holograms embossed on foil cost around 0.01 euros per hologram, with a total market volume of approximately 1 billion euros per year [5]. It is expected that the total security hologram market will reach $5,233.2 million by 2023 [6]. Typically, security holograms contain a surface phase relief [7]. All security holograms have one common drawback. While it is possible to hinder or even prevent contact copying of a holographic relief [8], it is not possible to eliminate the possibility of holographic copying of a hologram. Holographic copying involves restoring the image from the hologram using a beam of coherent light and re-recording the hologram of the reconstructed image. A thin layer of metal can be deposited on a copy hologram with a phase relief, and then a metal master stamp can be created through electroplating using standard technology [7,9,10]. This master stamp can be used to replicate copies of the original security hologram using the embossing method.

This paper discusses an optical security element (OSE) that allows for mass replication, similar to security holograms, but eliminates the possibility of holographic copying. This is achieved by incorporating a phase relief into the proposed element, which is not a hologram but rather a rough surface area [11,12]. The relief of a rough surface is random, and an area with a sufficiently large number of irregularities is almost unique. This allows it to be used for security and authentication [13,14]. Identification of optical security elements OSEs containing random phase inhomogeneities using speckle patterns formed under coherent illumination was carried out in a number of articles, for example, in [13-17]. However, in the mentioned publications, similar speckle patterns were formed by the same OSE, since the copies of the OSE were not produced. In particular, in [13], the formation of a speckle pattern occurs as a result of the passage of a beam of coherent light through an optical element, called a token, containing a layer of epoxy with glass spheres. It seems to us that replication of a token, which could form a similar speckle pattern, is fundamentally impossible. In other words, a token with these characteristics exists in only one copy and can be used, for example, as an ID card or a police badge. However, we were interested in OSEs that allow mass replication, similar to the replication of conventional security holograms. To verify the authenticity of such OSEs, it is necessary that OSEs obtained by embossing from a certain master stamp have reliably diagnosable differences from possible fakes. In our work, speckle patterns formed both by several replicas from the same rough surface and by replicas from different rough surfaces were compared. The speckle pattern is a granular structure in the intensity distribution of coherent light, resulting from the interference of coherent waves with a random phase scattered by different points on the rough surface [18-22]. The term "rough surface" as used herein means a surface containing random irregularities with an average size in the order of the wavelength of visible light. The suggestions made above have been experimentally verified.

## 2. Experimental

Moving on to the experimental part, let's start by discussing the process of obtaining a rough area of the OSE. In everyday life, a significant amount of objects have a matte surface. At first glance, it seems that in order to obtain a rough OSE area, it is enough to take a replica from an area of any available rough surface. However, in order to form speckle patterns with a high level of correlation, it was necessary to obtain several high-precision centimeter-sized copies from one rough area. Speckle images have been found to be very sensitive to replicating errors. A rough estimate of the permissible error gives about $\lambda/8$ (here $\lambda$ is the laser light wavelength) in the direction perpendicular to the surface of the OSE. The error in directions along the surface is much less critical. Reducing the influence of these errors was achieved by choosing rough surfaces with sufficiently smooth and not too small irregularities. For this purpose, plates made of Teflon (polytetrafluoroethylene) with a surface roughness of about 500 nm, achieved by grinding and polishing, were chosen as samples from which replicas were taken. Teflon was selected for its combination of sufficient hardness and reduced adhesion when detaching replicas from its surface. Thin layers of a photopolymer, which could be cured by UV radiation, were deposited on the surface of the substrates (Fig. 1). The substrates used were glass and quartz plates. The process involved aligning the substrate with the applied polymer layer with the surface of the Teflon plate under an optical microscope. Subsequently, the photopolymer was cured through UV radiation from the side of the substrate. The thickness of the photopolymer layer after alignment was controlled with the use of solid limiters located along the perimeter of the area of the copied surface and was approximately 50 μm. The transverse dimensions of the working field of the replica ranged from 5x5 to 10x10 mm$^2$. To increase the reflectivity of the replicas, a layer of 100 nm thick ruthenium was deposited on the surface of a random relief by magnetron sputtering. The transmittance of such layer was $2.9 \times 10^{-3}$.

The authentication of the produced OSE is demonstrated below. If you have a genuine OSE and an element whose authenticity needs to be verified, you must first illuminate an area of their rough surface with a beam of coherent light. Reflection from a random three-dimensional relief should produce a speckle pattern. If the copies of the rough relief area are identical, then the speckle patterns of the light scattered by them should also be identical when the same optical scheme is used. In practice, the relief profile of different specimens will have some differences, and there are always inaccuracies in the installation of the element in the optical scheme, instrumental noise, etc. Therefore, the corresponding speckle patterns will be somewhat different. The difference between two speckle patterns can be assessed by calculating the cross-correlation coefficient [23] of their digital images. For identical images, the correlation is close to 1, while for uncorrelated images, it is close to zero. The threshold value is set to ensure reliable identification in the presence of real inaccuracies.

The optical scheme of the experiment of authenticity identification of OSEs is shown in Fig. 2. The presented optical scheme was assembled in two versions: on a conventional optical table and on an optical bench. The first option was more convenient for research and initial adjustment of the optical scheme. The second option (Fig. 3) made it possible to place optical components more compact, which was convenient when transporting and demonstrating the installation. In the experiment, a semiconductor laser diode with a light wavelength of 650 nm (denoted as Laser1 in Fig. 2) was used. The power of the laser beam could be adjusted in the range of 0.1-2 mW. The outgoing laser beam, with a diameter of approximately 1.5-2 mm, was expanded using a set of lenses to create a uniform spot of at least 10-15 mm. The size of the working area of the sample was limited by aperture 1 (Fig. 2). In the experimental setup on the optical table, a photo image recording system based on the ToupCam U3CMOS05100KPA digital camera (5 MP matrix, matrix size 5.70x4.28 mm and a resolution of 2560x1922 pixels) was used. The focal length of the lens was 75 mm. To control the video sensor parameters (resolution, exposure time, etc.) and save images, a digital camera was connected to a computer and standard software supplied with the photo equipment was used. In the compact setup on an optical bench, a budget Supereyes B011 USB digital microscope with a 5 MP sensor was used. A modified measurement program was used in this setup to automate the saving of a series of images on the computer with varying photography parameters. To facilitate quick installation or replacement of samples, a snap-in mechanism was created for interchangeable holders with glued OSE test samples. This mechanism made it possible to regulate the alignment of sample positions and the angular orientation of the sample plane. It was essential to create the same conditions for the formation of speckle patterns from corresponding areas on the surface of the samples. To align the plane of the samples, interchangeable holders were installed in an adjustable frame, a beam of an auxiliary alignment laser (Laser2 in Fig. 2.) was directed to the area of the substrate of the samples, and the direction of the reflected beam was adjusted with the help of adjusting screws according to the position of the laser spot on the screen (Fig. 2). At the first stage, a rough alignment of the working areas of the surface of the samples was carried out visually according to the scratches seen on their image on a computer monitor, obtained using a digital camera. A more accurate alignment was performed when processing speckle patterns in the program for calculating the correlation function and searching for its maximum. Fig. 4 shows photos of the working areas (~9 x 7 mm) of the rough surface of the samples, taken under incoherent illumination with white light. Here, pictures (a) and (b) show the different replicas from the rough surface of Teflon plate #1 (replicas 1a & 1b), while pictures (c) and (d) show the different replicas from the plate #2 (replicas 2c & 2d). The speckle pattern was recorded with a maximum resolution of 2560x1922 pixels. The average speckle size was controlled by the size of the iris diaphragm placed in front of the camera lens (Aperture2 in Fig. 2). If speckles are observed in a certain plane using a lens with an aperture of diameter D, then the average speckle diameter in the observation plane will be equal to $1.22\lambda z/D$, where z is the distance from the lens to the viewing plane (section 1.4 in [20]). The highest correlation values were obtained with average linear speckle sizes of 5-10 pixels.

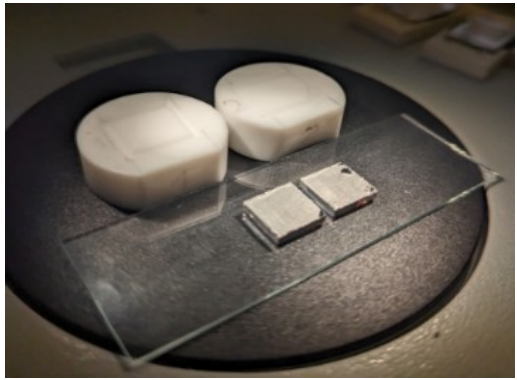

Fig. 1. The Two Teflon plates, labeled as #1 and #2, along with corresponding polymer replicas of the central part of their surface on quartz substrates.

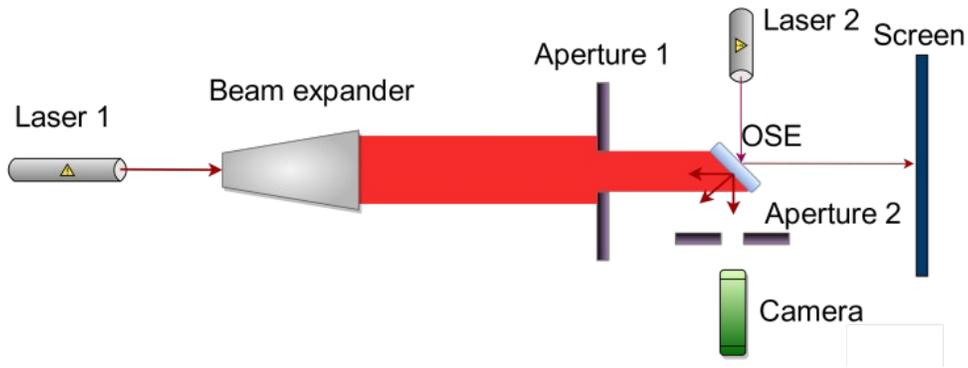

Fig. 2. The optical scheme of the experiment to identify the authenticity of OSEs.

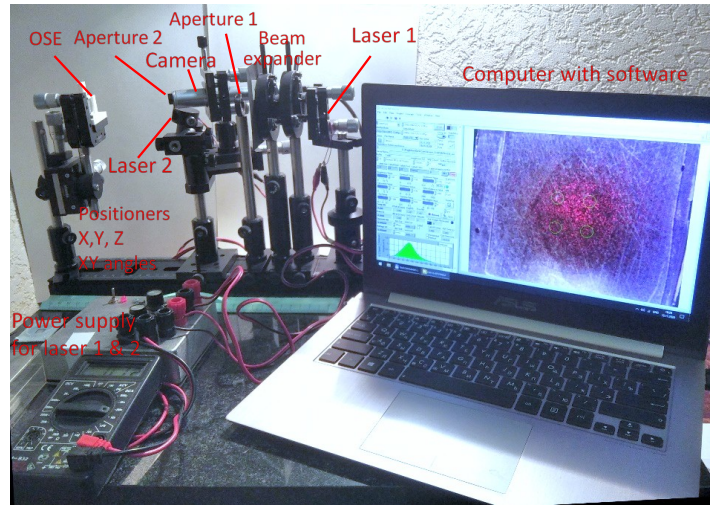

Fig. 3. Photo of the experimental setup. In the option presented here, the Laser 2 beam falls on the opposite side of the OE compared to Fig. 2.

3. **Results and discussion**

Fig. 5 shows fragments (1000x1000 pixels) of images of speckle patterns formed by two different replicas from the rough surface of Teflon plate #1 (replicas 1a & 1b, Fig. 5 (a) and 5 (b)) and plate #2 (replicas 2c & 2d, (Fig. 5(c) and 5(d)). To compare speckle patterns, a program was written in Python using the standard NumPy and SciPy libraries for working with data arrays. The program utilizes an optimized version of the 2DCC (2D cross-correlation) algorithm [14-16]. The 2DCC algorithm calculates the cross-correlation of two images by moving one image on top of the other and calculating the dot product of the two matrices. The shift at which the correlation is maximum is considered correct. The correlation of an arbitrary image with itself will always be equal to 1. Since the speckle patterns of different samples can be translated and rotated in the XY-plane relative to each other, in addition to the standard matrix shifts for 2DCC, the calculations were also carried out at different rotations. To visually represent the presence or absence of a correlation between two speckle patterns, a heat map was used. The heat map shows the dependence of the correlation coefficient on the image shift relative to each other along the X and Y axes. Fig. 6 shows heat maps associated with speckle patterns generated by replicas from the same rough surface area (a) and replicas from different rough surface areas (b). In heat maps (a) and (b), different color scales are used due to a significant difference in the ranges of changes in the correlation coefficients. For speckle patterns generated by

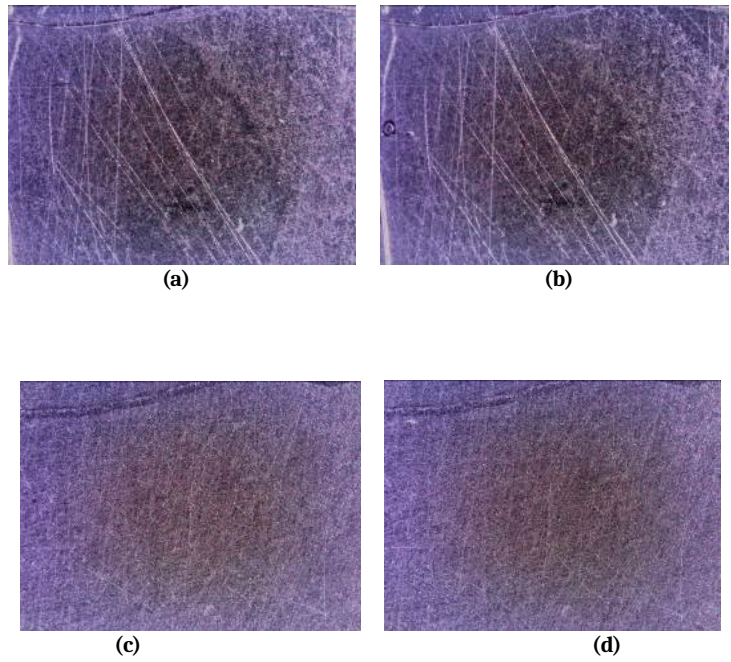

Fig. 4. Photos of the rough surface of the samples, taken under incoherent illumination.

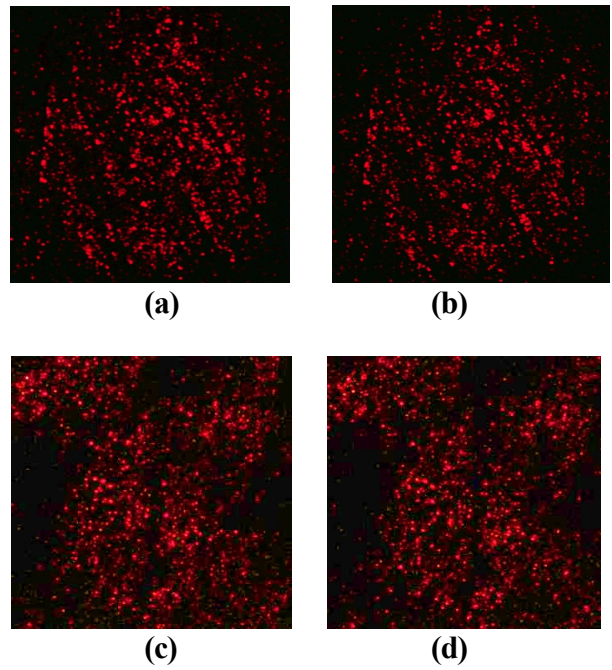

Fig. 5 Photos of the speckle patterns formed by two different replicas from the rough surface of Teflon plate #1 ((a) and (b)) of and from the plate #2 ((c) and (d)) (in these illustrations, data parameters have been changed to improve print quality).

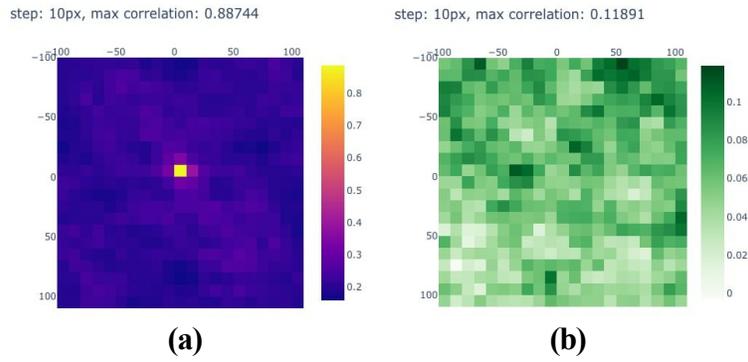

Fig. 6. Heat maps of the correlation coefficient between speckle patterns created by replicas from the same surface (a) and from different surfaces (b). In heat maps (a) and (b), different color scales are used due to a significant difference in the ranges of changes in the correlation coefficients.

Table1. Cross-correlation coefficients

|    | 1a         | 1b         | 2c         | 2d         |
|----|------------|------------|------------|------------|
| 1a | 1.0        | 0.9±0.05   | <0.06      | <0.08      |
| 1b | 0.9±0.05   | 1.0        | <0.05      | <0.06      |
| 2c | <0.06      | <0.05      | 1.0        | 0.85±0.05  |
| 2d | <0.08      | <0.06      | 0.85±0.05  | 1.0        |

replicas of the same surface (Fig. 6a), a single maximum with an amplitude close to 1 is observed on the heat map. Fig. 6 clearly shows the qualitative difference between the heat map in the presence of a real correlation and the heat map when the correlation is random. The values of the cross-correlation coefficients obtained as a result of processing the images of speckle patterns are shown in Table 1. It can be seen that the correlation between speckle patterns corresponding to replicas from the same rough surface significantly exceeds the similar correlation corresponding to replicas from different surfaces. This proves the efficiency of the used identification method.

At the first stage, comparison of speckle patterns required additional adjustment of the angular orientation of the sample plane and alignment of the positions of the OSE samples. Subsequently, to facilitate quick installation or replacement of samples, a mechanism for snapping interchangeable holders with OSE samples was manufactured on a 3D printer. This mechanism made it possible to automatically maintain the angular orientation of the samples and match the position of the samples. Computer processing determines the maximum correlation for various shifts of speckle patterns relative to each other, which actually means their exact alignment.

Brief information in [14] indicates that mechanical damage to the OSE does not have a significant effect on its identification from the corresponding speckle pattern. Moreover, the identification process is well protected not only from damage, but also from the complete loss of part of the surface of the random relief. Our data suggest that the loss of part of the sample surface degrades the signal-to-noise ratio, but allows identification up to certain limits.

Let's discuss possible variants of the OSE structure and methods for their identification. In addition to the microstructure of a rough surface, a three-dimensional OSE relief can contain details that are visually observable during the identification process. The three-dimensional relief of the OSE may also contain areas that are a hologram, which is additional evidence of the full compatibility of the proposed technology with the known technology of security holograms. Authentication compares the verifiable OSE with a known genuine OSE. Particularly attractive is the use of an OSE containing simultaneously a three-dimensional relief with visually distinguishable details and areas with a random microrelief. A visually distinguishable detail may directly contain a random microrelief. Primary identification of the authenticity of an OSE with visually distinguishable details can be done initially with the naked eye and then with an optical microscope. To further identify the authenticity of an OSE, the procedure described above for comparing speckle patterns between verifiable and genuine OSEs should be followed.

As already mentioned, holographic copying of a three-dimensional relief area of an OSE is practically impossible. However, it is possible to record a hologram of such relief area. In this scenario, rather than containing a three-dimensional relief, the OSE under examination would contain a hologram of the genuine OSE's three-dimensional relief. When this hologram is illuminated by a beam of coherent light corresponding to the reference wave, a three-dimensional image of the relief is restored. However, the image of any object reconstructed from a hologram can be distinguished from a real object, since the geometry of the image reconstructed from a hologram will depend on the wavelength and angle of incidence of the reconstructing beam. In order to conclude based on the identity of the speckle patterns about the authenticity of the OSE to be checked, it is also necessary to exclude the case when instead of an OSE containing a rough relief area, its holographic copy is used. To do this, the operation of illuminating the test and genuine OSE with coherent light beams and comparing speckle patterns is repeated several times, while the radiation wavelength and/or the angle of incidence of the illuminating beam are changed in the same way. Then, if the object under test is a genuine OSE, the corresponding speckle patterns will remain identical, and if the object under test is a hologram, the corresponding speckle patterns will be different from each other.

Let's talk about the mass production of the proposed OSEs. The original three-dimensional relief containing details accessible to visual observation is obtained by known methods, for example, by engraving. In the resulting relief, the surface of details accessible to visual observation usually contains random microroughnesses with a characteristic size comparable to the wavelength of the visible light. Thus, the necessary relief roughness can be formed in a natural way, without the use of specially applied operations. Next, a polymer replica is taken off the surface of the original relief. A thin metal coating is applied to the replica by vacuum deposition or chemical deposition from a solution. A layer of nickel or its alloys with a thickness of 50-100 μm is applied to the obtained metallized surface by electroforming. As a result of the process, a metal master stamp is formed, which is mounted on the rotor of the OSEs embossing machine. Thus, for industrial replication of OSEs, it is proposed to use the technology of replication of security holograms [7,9,10], which differs from that used in this study. Therefore, studies of the actual manufacturing accuracy of OSEs should be carried out using industrial equipment for reproducing security holograms. We hope that this accuracy will be sufficient to reliably identify OSEs. Otherwise, it may be possible to use nanoimprint lithography methods, which have been rapidly developing since the 90s [27,28].

To protect the proposed OSEs from contact copying, the same methods that have been developed for traditional security holograms can be used. Access to the phase relief is protected by a special transparent layer. The most reliable way is to glue OSEs onto the protected object with the relief facing down, and the adhesive layer is applied directly to the relief. In this case, the hardened glue does not allow the hologram to be separated from the object without destruction. Consequently, contact copying of OSEs becomes almost impossible. Holographic copying of a random relief area is impossible, since it is not a hologram. Photographic copying of a random relief is also impossible due to its three-dimensionality, since the structure of speckle patterns significantly depends on the phase ratio of partial waves scattered by relief inhomogeneities.

4. **Conclusion**

Laboratory samples of OSEs were produced by taking polymer replicas from the rough surface of Teflon plates with submicron precision. Identification of the authenticity of the OSEs was carried out by comparing speckle patterns formed by laser light scattered off different replicas. Speckle patterns were recorded with a digital camera connected to a computer. As a result of computer processing, their cross-correlation was calculated. The experimental results show that this method allows for reliable determination of authenticity of the tested OSEs.

## CRediT authorship contribution statement

**A.M. Smolovich:** Conceptualization, Methodology, Investigation, Writing - Original Draft, Writing - Review & Editing. **A.V. Frolov:** Methodology, Validation, Investigation, Writing - Review & Editing. **L.D. Klebanov:** Investigation. **I.D. Laktaev:** Investigation. **A.P. Orlov:** Investigation, Formal analysis, Resources, Data Curation, Software, Methodology, Validation, Visualization, Writing - Original Draft, Writing - Review & Editing. **P.A. Smolovich:** Software, Formal analysis. **O.V. Butov:** Supervision, Writing – review & editing.

## Declaration of competing interest

The authors declare that they have no known competing financial interests or personal relationships that could have appeared to influence the work reported in this paper.

## Data availability

No data were generated or analyzed in the presented research.

## Acknowledgments

The work has no special funding. The work was carried out within the framework of the State task of the Kotelnikov IRE RAS. The authors thank A.A. Zlatopolsky for useful discussions of image processing methods, and A.S. Ilyin for consultations on the technology of taking replicas and for applying a reflective coating to the replicas.